\documentclass[fleqn,usenatbib]{mnras}

\usepackage{newtxtext,newtxmath}

\usepackage[T1]{fontenc}

\DeclareRobustCommand{\VAN}[3]{#2}
\let\VANthebibliography\thebibliography
\def\thebibliography{\DeclareRobustCommand{\VAN}[3]{##3}\VANthebibliography}

\usepackage{graphicx}	
\usepackage{amsmath}	

\usepackage{amssymb}	
\usepackage{ulem}
\usepackage[dvipsnames]{xcolor}

\newcommand{\msun}{M$_{\odot}$}

\title[Merging Black Holes in Dwarf Galaxies]{Merging Black Holes in Dwarf Galaxies: Calculating Binary Black Hole Coalescence Timescales from Simulations for LISA Detection}

\author[V. I. De Cun et al.]{
Victoria I. De Cun,$^{1,2,3}$
Jillian M. Bellovary,$^{4,5,3}$
Michael L. Katz$^{6}$
\\
$^{1}$Department of Astronomy, New Mexico State University, MSC 4500, PO BOX 30001, Las Cruces, NM 88003, U.S.A.\\
$^{2}$Department of Physics, Florida International University, 11200 SW 8th Street, Miami, FL 33199, U.S.A.\\
$^{3}$Department of Astrophysics, American Museum of Natural History, Central Park West at 79th Street, New York, NY 10024, U.S.A.\\
$^{4}$Department of Physics, Queensborough Community College, City University of New York, 222-05 56th Avenue, Bayside, NY 11364, U.S.A.\\
$^{5}$Department of Astrophysics, CUNY Graduate Center, 365 5th Ave, New York, NY 10016, U.S.A.\\
$^{6}$Max-Planck-Institut f\"ur Gravitationsphysik, Albert-Einstein-Institut, Am M\"uhlenberg 1, 14476 Potsdam-Golm, Germany\\
}
\date{Accepted XXX. Received YYY; in original form ZZZ}

\pubyear{2021}

\begin{document}
\label{firstpage}
\pagerange{\pageref{firstpage}--\pageref{lastpage}}
\maketitle

\begin{abstract}
Supermassive black holes (SMBHs) merging in dwarf galaxies will be detectable by the Laser Interferometer Space Antenna (LISA) in the mid-2030s. Previous cosmological hydrodynamic simulations have shown the prediction of massive black holes merging in dwarf galaxies, but these simulations are limited by their resolution and cannot follow black hole pairs all the way to coalescence. We calculate the delay time between black hole pairing and merger based on the properties of the black holes and their host galaxies, and use these properties to calculate gravitational wave strains for eleven different binary black holes that merge inside dwarf galaxies from eight cosmological simulations. This delay time calculation accounts for dynamical friction due to gas and stars, loss-cone scattering, and hardening of the binary due to gravitational radiation. Out of the eleven black hole mergers in the simulations, five black hole pairs will merge within 0.8 - 8 Gyr of forming a close pair and could be observed by LISA, and the remaining six are unresolved due to resolution limitations of the simulation. As all five of the resolved close pairs merge within a Hubble time, we make the broad estimate that close SMBH pairs in dwarf galaxies will merge and be detectable by LISA, but this estimate depends on either the presence of gas during orbital decay or a solution to the dynamical buoyancy problem in cored potentials.
\end{abstract}




\section{Introduction}

Dwarf galaxies are the most abundant and common type of galaxy in the universe, raising the question: are their black hole merger events just as common? Since dwarf galaxies merge, it raises the possibility that their black holes may also merge. Several studies have revealed evidence for the existence of supermassive black holes (SMBHs) in dwarf galaxies \citep{Reines2013, Moran2014, Satyapal2014, Lemons2015, Sartori2015, Pardo2016, Molina2021}, which are summarized here. 

Below, we review Active Galactic Nuclei (AGN) occupation fractions that were determined from previous studies. These studies support that AGNs exist in dwarf galaxies, which can likely suggest that these AGNs are inactive/active black holes within dwarf galaxies.
In \cite{Reines2013}, the AGN occupation fraction of optically selected active black holes in their parent sample of dwarf galaxies is $\sim0.5\%$ (136/25974). However, they state it is difficult to obtain the true occupation fraction and black hole mass function in this low-mass regime.  \cite{Moran2014} examines the ${{M}_{\star }}=(4-10)\times {{10}^{9}}$ ${{M}_{\odot }}$ range (that includes 75\% of the AGN host galaxies in their survey), and obtains an occupation fraction of 2.7\%. \cite{Satyapal2014} uses the all-sky Wide-Field Infrared Survey Explorer (WISE) survey and the bulge-to-disk decomposition from SDSS Data Release 7. They report their discovery of a population of local ($z<0.3$) bulgeless disk galaxies with extremely red mid-infrared colors which suggest a dominant AGN, despite having no optical AGN signatures in SDSS spectra. Their study finds that in the dwarf galaxy regime the AGN fraction is less than 2\%. In \cite{Lemons2015}, they present a sample of hard X-ray-selected candidate black holes in 19 dwarf galaxies. 8 of the 19 dwarf galaxies in the sample have X-ray detections reported for the first time, which suggest an AGN fraction upper limit of 42\%. \cite{Sartori2015} investigated AGN activity in low-mass galaxies and their interaction with their host galaxies. They identified 336 AGN candidates from a parent sample of $\sim 48,000$ nearby low-mass galaxies, giving an AGN fraction of $\sim 0.7\%$. In \cite{Pardo2016}, they present a sample of accreting SMBH in dwarf galaxies at $z<1$. From their sample of 605 dwarf galaxies, 10 exhibit X-ray emission consistent with that arising from AGN activity. They find that the AGN fraction ranges between $\sim 0.6\% - 3\%$. \citet{Polimera22} use local dwarfs with strong emission lines to determine a larger AGN fraction of 3-16\%.  Finally, \citet{Molina2021} search for AGN in dwarf galaxies using [Fe X]$\lambda 6374$ coronal line emission from SDSS spectra, and find 81/46530 dwarf galaxy AGN candidates.   This occupation fraction (0.17\%) is an upper limit, however, because it could also include tidal disruption events (which indicate the presence of an SMBH regardless).

It is clear that dwarf galaxies have AGN, but what about the black holes that are not active? With only a fraction of existing black holes accreting, it is difficult to determine the actual occupation fraction of black holes within dwarf galaxies. In \cite{Greene2020}'s Figure 5, they show predictions for the occupation fraction from recent models. From the three models depicted in the figure, the occupation fraction at $10^{9} M_{\star}$ ranges from 0\% to 100\%. Observationally,  \cite{Nguyen2019} infers an occupation fraction >50\% from dynamical measurements, and X-ray surveys of galaxies $M_{\star} = 10^{9} - 10^{10} M_{\odot}$ suggest a lower limit of 20\% on the occupation fraction \citep{Miller2015, She2017a}. 
The papers described above reveal that dwarf galaxies can not only host active black holes, but that potentially many more dwarf galaxies can host non-accreting black holes. The large uncertainties in the occupation fraction highlight the need to facilitate discoveries of more black holes in dwarf galaxies.

A new instrument for detecting gravitational waves is arriving with the construction of the Laser Interferometer Space Antenna (LISA), which will operate at a low frequency range between $\sim 10^{-4} - 10^{-1}$ Hz. This observatory will allow the possible detection of supermassive black hole (SMBH) merger events, like those suspected to be in the centers of most galaxies, including dwarf galaxies \citep{LISA2017}. The Laser Interferometer Gravitational-Wave Observatory (LIGO), operating at a frequency range of $\sim 10^{1} - 10^{4}$ Hz, has produced data on stellar-origin black holes, however, LISA will be able to detect gravitational waves that have much lower frequencies than LIGO. \textit{The data we have collected and present throughout this paper are used to make predictions for SMBH mergers in dwarf galaxies which can be detected by LISA.}

\cite{Bellovary2019} performs a study of several high-resolution, cosmological, zoom-in simulations that focus on dwarf galaxies that host massive black holes at $z=0$. They report that larger/more massive dwarf galaxies are more likely to host massive black holes rather than those of lower mass, and about 50\% of the massive black holes are not located in the centers of their galaxies, but wandering within a few kpc of the center. Furthermore, \cite{Bellovary2019} predicts that 11 binary black hole mergers are in the frequency range of detection by LISA. We will examine those 11 binary black hole mergers within this paper.

 Using data from the Illustris cosmological simulation, \citet{Katz2019} performs an analysis of LISA detection rates along with the characterization of the merging massive black hole population (with total mass between $\sim 10^{5} - 10^{10}$). However, this work does not extend to the dwarf galaxy regime. Our research uses the method from \citet{Katz2019} using the \citet{DA17} (DA17) orbital decay model in addition to models of orbital decay due to gas with our sets of high-resolution simulations (which include dwarf galaxies with black holes);  the simulations are described in detail in \cite{Bellovary2019}. Both DA17 and \citet{Katz2019}  perform this analysis by calculating delay times and merger timescales. With these methods, we can calculate realistic merger timescales for the events reported in \cite{Bellovary2019} and analyze their gravitational wave signals.

Our goal is to determine how long it takes for close black hole pairs to merge in dwarf galaxies, and whether LISA can detect these possible merger events. We use cosmological simulations including  formation and mergers of  black holes, and use analytic models to calculate delay timescales of these black hole merger events. An important aspect of this research is accounting for large scale decay, dynamical friction, loss-cone scattering, and gravitational wave hardening, all which lead to coalescence. In this paper we account for these essential stages, which are lacking in the previous cosmological simulations.

\section{Methods}

To predict gravitational wave signals from merging black holes in dwarf galaxies, we start with zoom-in cosmological hydrodynamic simulations run using the smoothed particle hydrodynamics ChaNGa code \citep{Menon15}. We then use an analytic prescription based on work by DA17 and \citet{Katz2019} to estimate full merger timescales for black holes in dwarf galaxies.

\subsection{Cosmological Simulations}
\label{sec:Cosmo Sims}

We use two sets of high-resolution simulations which include dwarf galaxies with black holes, which are described in detail in \citet{Bellovary2019}, and we summarize the vital details here. One suite, known as the ``MARVEL-ous Dwarfs'' consists of four volumes containing dwarf galaxies, with a force softening resolution of 60 pc, dark matter particle masses of 6660\msun,  gas particle mass of 1410\msun, and star particle mass of 422\msun. For the other suite, known as the ``DC Justice League,'' each simulation consists of one large Milky Way-like galaxy surrounded by many dwarf galaxies, some of which host black holes as well. These simulations have a force softening resolution of 170 pc, dark matter particle masses of $4.2 \times 10^4$\msun, while gas particles have a mass of $2.7 \times 10^4$\msun ~and star particles 8000\msun. 

All simulations use the same physical recipes, including molecular hydrogen-based star formation \citep{Christensen12}, blastwave supernova feedback \citep{Stinson06}, and a cosmic UV background \citep{Haardt12}. This set of simulations is the highest-resolution set of simulations which follow black holes in dwarf galaxies from the epoch of black hole formation to the present day.

Black holes form self-consistently in our simulations from gas particles which are  collapsing, overdense (3000 cm$^{-3}$ for the MARVEL-ous Dwarfs and $1.5 \times 10^4$ cm$^{-3}$ for the DC Justice Leauge),  have low metallicity ($Z < 10^{-4}$), and low molecular hydrogen fractions ($f_{H_2} < 10^{-4}$).  These criteria most closely mimic the ``direct collapse'' model of black hole formation, but they are also applicable to other models such as Population III stars or some frameworks of collapsing clusters.  The former model is expected to occur in the most overdense regions, which are not expected to be the locations of dwarf galaxies.  The latter models may be more globally applicable, but result in smaller mass black holes.  However these objects may undergo rapid accretion, growing to intermediate sizes very quickly \citep{Volonteri05}.  Our simulations aim to form seed black holes using physical conditions that are broadly consistent with any of these mechanisms.  In these simulations, the black hole seed mass is 50,000\msun, though \citet{Bellovary2019} explains how seeds can actually form above or below this value.  (See their Figure 1 for a depiction of all black hole seed masses in these simulations.) The efficiency of seed formation likely varies depending on the mechanism, with direct collapse being the most stringent.  However, even if lower mass seeds are more plentiful, rapid accretion events may not be.  The occupation fraction of seeds in dwarfs depends on these and other factors, which are all unknown.  We thus postulate that our seed formation recipe may provide a lower limit to the occupation fraction, due to the somewhat strict constraint on molecular hydrogen fraction.

Particularly relevant to this work is the implementation of dynamical friction on black hole particles via a sub-grid model as described by \citet{Tremmel15}.  This model is based on the Chandrasekhar formula \citep{Chandrasekhar43,Binney08} and estimates dynamical friction on scales smaller than the softening length of the simulation.  With this formalism, we can accurately track the trajectories of black holes as their host galaxies merge and they eventually form close pairs, without relying on artificially pinning them to the centers of their hosts.  Due to this model, we can realistically follow black hole orbital evolution until they reach and maintain a distance of $\sim$a hundred parsecs, mimicking the close pair regime.  


As reported in \citet{Bellovary2019}, our simulated dwarf galaxies themselves reflect the properties of observed local dwarfs.  Specifically, they lie on the observed stellar mass - halo mass relation and have star formation histories which bracket the diversity of dwarfs in the local group.  In \citet{Bellovary2019} work, they identified several black hole mergers in dwarf galaxies.  However, their definition of a merger actually is a close pairing, because the spatial scales of mergers are not resolved.  They allow black holes to coalesce if they become close together in space ($< 2$ force softening lengths) and have low relative velocities (fulfilling the criterion $\frac{1}{2}\Delta {\vec{ \rm v}} < \Delta {\vec{ \rm a}} \cdot \Delta {\vec{ \rm r}}$, where $\Delta {\vec{ \rm v}},  \Delta {\vec{ \rm a}}$ and  $\Delta {\vec{ \rm r}}$ represent the relative velocity, acceleration, and radius vectors of the two SMBHs respectively). The purpose of this work is to fill in the gap between the resolution limit and actual black hole merging.


\subsection{Delay Time Calculation}
\label{sec:DA17} 

Black hole merger events occur in several stages: large scale decay, dynamical friction, hardening, and finally, coalescence. These essential stages are explained below.

Large scale orbital decay, or large scale delay, has the largest orbit and is a gradual decrease of the distance between the two orbiting black holes at their closest approach over many orbital periods. These orbits do not decay without some friction-like mechanism that transfers energy from the orbital motion, which, in this case, would be due to dynamical friction. During the large scale orbital decay stage, the galaxies and black holes are still separated at a distance of $\sim 10-100 \;$ kpc. This separation gradually decreases over many periods as they pass near their closest approach. This decay is due to dynamical friction governed by a distribution function that is assumed to be Maxwellian \citep{Katz2019}. This process occurs until the binary reaches a separation of $\sim$1 kpc. At this stage, the two black holes are close enough that their environments are influenced by each other. This process changes the distribution function to a more complicated form given in Equation 20 in DA17. Once the binary reaches separations of $\sim$pc, hardening takes over as the dominant mechanism for orbital decay. Hardening occurs when stars in fast orbits interact with the bound pair of SMBHs and remove energy from the binary system. This is also referred to as "loss-cone scattering" \citep{Frank1976, Lightman1977}. At a separation smaller than the hardening separation (see Equation 26 in \cite{Vasiliev2015}), gravitational radiation takes over as the dominant decay mechanism and drives the binary to coalescence. While previous simulations have shown promising results for black hole merger events, they do not account for these essential stages that lead to coalescence.

The DA17 model assumes that the mergers occur in gas-poor ellipticals, and that stars dominate the dynamical friction.  We argue that the collisional aspect of this model applies for dwarf galaxies as well.  The relevant quantities that determine inspiral times are valid at any scales (e.g. density profile slope $\gamma$ and stellar velocity dispersion $\sigma$), and in fact the flat density profiles expected by DA17 for giant ellipticals are not drastically different from the cored ones in dwarf galaxies.  However, our galaxies also include gas, which may contribute significantly to the dynamical friction of the SMBHs. We therefore include both collisionless and gas-dynamical processes as we calculate coalescence times.

Large scale decay allows for separate galaxies (each hosting black holes) to merge into one galaxy, but the black holes are still separated. The large scale decay stage is the first step in determining a delay timescale for these black hole merger events, and \cite{Katz2019} performs this calculation because the galaxies are still separate when the calculation begins. However, we are able to follow two galaxies with two black holes until the galaxies merge together, but the black holes are still separate. Thus, we resolve full galaxies merging when black holes are in separate galaxies, as the distances in Table \ref{tab:galaxy properties} show that the black holes are already in the same galaxy and near each other when we begin our calculation. We calculate the large scale decay timescales to be between $10^{6}$ and $10^{7}$ years, and are always orders of magnitude shorter than the hardening timescale. Large scale decay does not affect delay timescale calculations, thus, we do not include this calculation for our simulations in the total timescale calculation.

The next stage is dynamical friction, where the timescale for the binary to decay to a shorter separation is given by a variation of the Chandrasekhar formula (DA17) , 

\begin{equation}
\begin{split}
    T_{\rm OD, star} = 1.5 \times 10^{7} \frac{[\rm ln \Lambda \alpha \; + \; \beta \; + \; \delta]^{-1}}{(3/2 \; - \; \gamma)(3 \; - \; \gamma)}(\chi^{\gamma - 3/2} \; - \; 1) \\ \; \times \; (\frac{M}{3 \times 10^{9}})^{1/2} (\frac{m}{10^{8}M_\odot})^{-1} (\frac{r_{\rm infl}}{300 \; pc})^{3/2} \; yr,
    \label{eq:Tbulletbare}
\end{split}
\end{equation}

 \noindent{where $M$, (secondary, $m$) denotes the larger (smaller) MBH,  $r_{\rm infl}$ is the influence radius of the primary MBH approximated from \cite{Merritt}, $\rm ln \Lambda$ is the Coulomb logarithm,  $\gamma$ is the power law exponent in the stellar density profile, $\chi = a_{h}/r_{\rm infl}$, \; and $\alpha, \beta,$ and $\delta$ are functions calculated from Equations 17-19 in DA17.\footnote{The dynamical friction decay timescale is not greatly affected by the orbital eccentricity, which, in these equations, are assumed to be a circular orbit $\xi = 1$.}}  
 
 This formula is applicable in the cases of infalling objects, but prior works have noted that in the case of cored density profiles (as our dwarfs exhibit here) core-stalling will occur and dynamical friction is essentially zero \citep{Read06, Petts15}.  This effect is due to an effective dynamical buoyancy which occurs when massive objects reach the center of a cored object, caused by the drastic change in the radial gradient of the density distribution \citep{Banik22}.  Once the black holes arrive in the cored region of the galaxy, they may in fact never find each other, and mergers will not occur.  However, these prior works (whether analytic or numerical) are based on assumptions such as spherical symmetry and functional density distributions, and these assumptions are less valid in the case of a dynamically evolving galaxy.  In an attempt to bracket the ranges of timescales of orbital decay, we implement the result of \citet{Kaur21}  who find that in some cases, higher-order non-corotating resonances cause black holes in cored potentials to experience a dynamical friction force of 10\% that of Chandrasekhar.  We thus multiply our calculations in Equation \ref{eq:Tbulletbare} by a factor of 10 to represent the pessimistic case.  The calculated timescales are shown in Table \ref{tab:all}.

We additionally calculate the role of gas dynamical friction using the methodology developed in \citet{Chapon13}, which uses equations from \citet{Ostriker99} to calculate the Bondi drag of a black hole moving through a gaseous medium.  The wake which trails the black hole is most effective at changing its acceleration at a Mach number of $\mathcal{M} =1-3$.  The dynamical friction force is

\begin{equation}
F_{\rm DF, gas} = 4\pi\rho(GM_{BH}/c_s)^2  f_{\rm gas}
 \label{eq:gasDF}
\end{equation}

where we measure the local gas density $\rho$ and the local sound speed $c_s$ within a 1 kpc radius around each black hole. The factor $f_{\rm gas}$ in the supersonic case (which applies here) is given by 

\begin{equation}
f_{\rm gas} =  \frac{1}{\mathcal{M}} \left [ 1/2  \ln (\mathcal{M}^2 - 1) + \ln \Lambda \right ]
 \label{eq:f_factor}
\end{equation}

where the Coulomb Logarithm $\Lambda$ is set to a value of 3 \citep{Chapon13}.  The orbital decay timescale is given by 

\begin{equation}
T_{\rm OD,gas} =  \frac{L}{\dot{L}} =  \frac{M_{\rm BH} v_{\rm circ}}{F_{\rm DF}}
 \label{eq:gas_orbitaldecay}
\end{equation}

Since the black holes are not literally in a bound orbit at this stage, rather than $v_{\rm circ}$ we use the greater velocity of the two black holes in the galaxy, which is always larger than $v_{\rm circ}$.  Similarly, for the Mach number we use the value of $\mathcal{M}$ for whichever black hole's local gas sound speed results in a longer calculated $T_{\rm OD,gas}$.  In general, the values of these quantities are within 50\% of each other for each black hole pair, so we make the conservative assumption and use the longer timescale.

It is likely that one of these processes (gaseous or stellar dynamics) will dominate over the other, considering the different physical scenarios involved.  To estimate the total orbital decay timescale due to dynamical friction, we take the minimum of the two timescales  $T_{\rm DF} = min(T_{\rm OD, star},T_{\rm gas, OD)}$. Table \ref{tab:all} presents the calculated timescales for both the stellar-dominated and gas-dominated processes, and shows that in two of the five cases the gas dynamical friction process is dominant, while in the other three stellar dynamics are likely more dominant.  However, in every case, the orbital decay phase is shorter than the final inspiral phase - the hardening of the binary.

The binary is evolved with Equation \ref{eq:Tbulletbare} down to the hardening radius, $a_{h}$, which is the distance between the black hole binary by gravitational waves. It is given by \citep{Merritt2013},

\begin{equation}
    a_{h} \approx 36 \frac{q}{(1 \; + \; q)^{2}} \frac{M \; + \; m}{3 \times 10^{9} M_\odot} (\frac{\sigma}{300 \; km s^{-1}})^{-2} \; pc,
    \label{eq:hardeningradius}
\end{equation}
\\
\noindent{where the mass ratio is denoted by $q$ ($q \leq 1$), and $\sigma$ is the three dimensional stellar velocity dispersion.}   

Hardening is the final stage in the DA17 model, and includes the effect of gravitational radiation. The timescale, from $a_{h}$ until coalescence, is given by \citep{Vasiliev2015},

\begin{equation}
\begin{split}
    T_{\rm h, GW} \approx 1.2 \times 10^{9} (\frac{r_{\rm infl}}{300 \; pc})^{\frac{10+4\psi}{5+\psi}} (\frac{M \; + \; m}{3 \times 10^{9} M_\odot})^{\frac{-5-3\psi}{5+\psi}} \\ \; \times \; \phi^{-\frac{4}{5+\psi}} (\frac{4q}{(1 \; + \; q)^{2}})^{\frac{3\psi-1}{5+\psi}} \; yr,
    \label{eq:ThGW}
\end{split}
\end{equation}

\noindent{where $\phi = 0.4$ and $\psi = 0.3$ are triaxial parameters estimated from Monte Carlo simulations in \cite{Vasiliev2015}. In Equation \ref{eq:ThGW}, the eccentricity factor is left out as it is unity because circularity is assumed. }

The above formalism neglects the presence of gas, which is though to accelerate merging black hole systems \citep[e.g.][]{Lodato09}.     Using Equation 3 from \citet{Bortolas21}, we calculate the hardening timescale due to gas as

\begin{equation}
    T_{\rm h, gas} = \frac{a} {\dot{a}} = \frac{m}{2.68 \dot{m}}
    \label{eq:harden_gas}
\end{equation}

where $m$ is the sum of the black hole masses and $\dot{m}$ is the sum of the accretion rates of each black hole.  We calculate an average accretion rate from the simulation (see \citet{Bellovary2019} for the precise accretion model) within a duration of $10^4$ years before the merger.  Again assuming one process (dynamics/gravitational waves vs gas) will be dominant, we take the minimum of the two  timescales to be the final hardening timescale  $T_{h} = min( T_{h, GW}, T_{h, gas} )$.  In four of the five cases, $T_{h, GW}$ is the shorter timescale, suggesting that in general gas dynamics are less important for binary hardening in dwarf galaxies, likely due to the lower accretion rates of the black holes.

The final coalescence timescale, $t_{\rm coal}$, based on the DA17 model, is given by,

\begin{equation}
    t_{\rm coal} = T_{DF} \; + \; T_{h}.
    \label{eq:tcoal}
\end{equation}

\begin{figure}

	\includegraphics[width=\columnwidth]{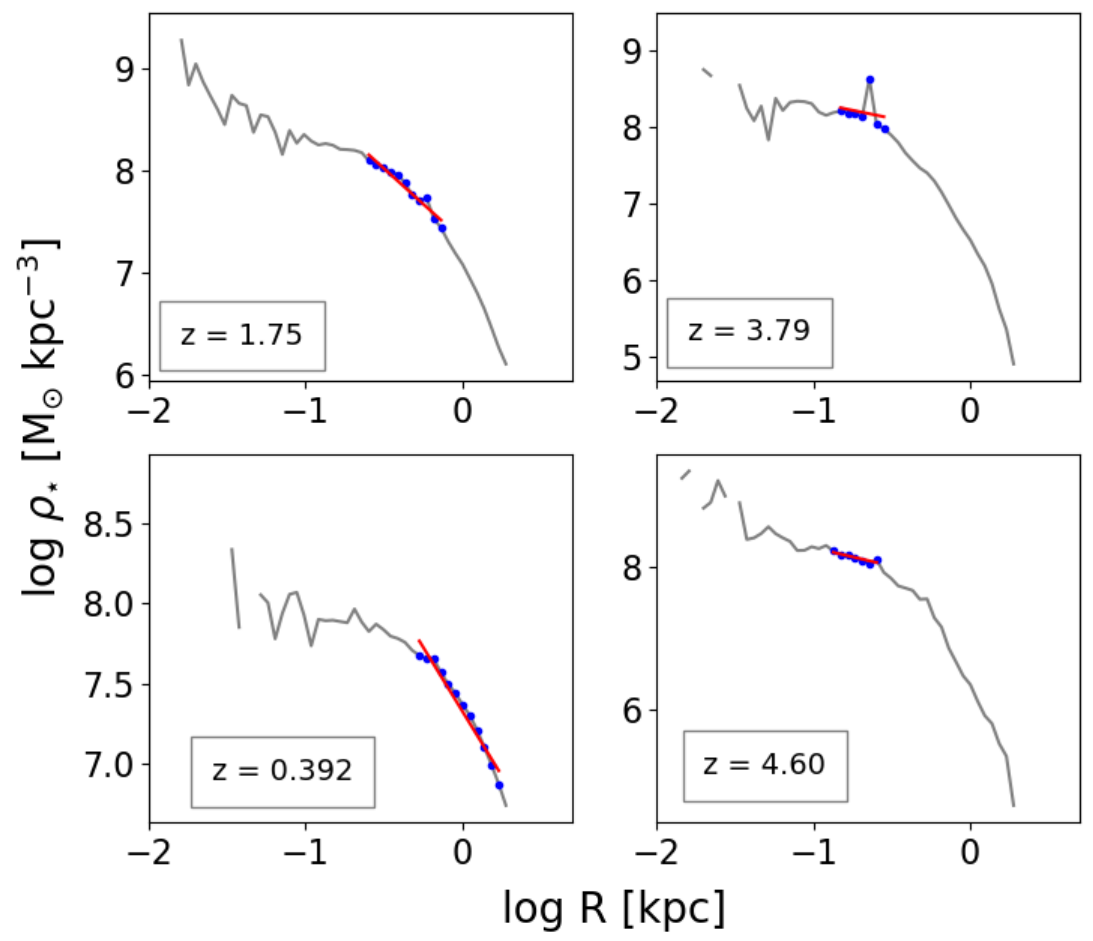}
    \caption{This figure displays the stellar density profiles of the galaxies hosting the black holes at the redshifts which the black hole mergers occur. The y-axis is the density, and the x-axis is the radius. The gray lines represent the stellar density of the whole galaxy, blue dots represent the stellar density at a specific range where the inner radius is set to four times the gravitational softening (170 pc) and outer radius is set to 2\% of the virial radius, and the red lines are fit to the blue points to measure slopes. }
    \label{fig:multipanelplot}
\end{figure}

 The properties of the galaxies which are used to calculate the timescales can be seen in Table \ref{tab:galaxy properties}. Figure \ref{fig:multipanelplot} displays the stellar density profiles of the galaxies hosting the black holes at the redshifts which the black hole mergers occur. The gray lines represent the stellar density of the whole galaxy, blue dots represent the stellar density at a specific range where the inner radius is set to four times the gravitational softening (170 pc) and outer radius is set to 2\% of the virial radius, and the red lines are fit to the blue points to measure slopes. The calculation for the velocity dispersion, gas density and sound speed uses a radius set to $1$ kpc around each black hole. 

\section{Data and Results}
The subsequent sub-sections, \ref{sec:Delay Time} and \ref{sec:Gravitational Wave Signals}, discuss the results of the delay time calculation for five of the 11 mergers, gravitational wave strain plot of the 11 mergers and a plot with all the merger events as a function of redshift and the combined black hole mass. Six mergers are unresolved because there were not enough particles in these simulations to robustly measure galaxy properties. The calculated results show that four of the five mergers will merge within a Hubble time and can be observed by LISA. Sub-section \ref{sec:Comparisons to Prior Work} compares our research to prior studies.

\subsection{Delay Time}
\label{sec:Delay Time}

Table \ref{tab:galaxy properties} provides the properties of the coalescing black holes as well as the properties of the galaxies which host each black hole merger. We then calculate the delay and coalescence timescales as described above, shown in Table \ref{tab:all}. Table \ref{tab:all} shows the delay times (in years) for each black hole merger. The delay timescale is the sum of dynamical friction (whether due to stars or gas) and hardening (whether due to gas or gravitational radiation and loss-cone scattering). Coalescence time is the age of the universe when black holes merge plus the delay timescale, indicating whether the signal would propagate to our location on Earth within a Hubble time.  

For all five mergers, the hardening timescale is 1-2 orders of magnitude larger than the orbital decay timescale.  In all but one case this is true whether we use the stellar- or gas-dynamical friction timescales, indicating the fairly robust result of hardening being the main bottleneck.  In four out of five cases, the hardening is dominated by stellar processes and gravitational waves rather than gas; however in the case where gas is dominant it speeds up this phase of the merger by a factor of 10, from $\sim 32$ Gyr to 3.1 Gyr.  We conclude that in a very gas-rich environment (where black holes accrete efficiently) coalescence can be sped up considerably, but this may not be a common occurrence.

\begin{table*}
    \centering
    \caption{This table provides the black hole (BH) and galaxy properties determined in order to calculate delay timescales.}
    \label{tab:galaxy properties}
    \begin{tabular}{c|p{17mm}|p{15mm}|p{15mm}|p{15mm}|p{15mm}|p{15mm}|c}
         \hline
         Redshift & Total BH Mass  $(M+m)$ $(M_\odot)$ & Mass Ratio $(\frac{m}{(M+m)})$ & Distance Between BH (kpc) & Stellar Velocity Dispersion (km s$^{-1}$) & Gas Density ($M_\odot$ kpc$^{-3}$)& Local Gas Mach \hspace{2mm} Number &  Stellar Mass ($M_\odot$) \\
         \hline
         9.42 & $4.66 \times 10^{6}$ & 0.292 & 1.32 & 24.3  &  $2.11  \times 10^{8}$    & 2.62 &$1.04 \times 10^{7}$ \\
         4.60 & $2.24 \times 10^{6}$ & 0.310 & 0.357 & 29.9 &   $2.42  \times 10^{7}$   &  4.23  &$7.71 \times 10^{7}$ \\
         3.79 & $8.22 \times 10^{4}$ & 0.479 & 0.141 & 18.7 &   $2.11  \times 10^{9}$   & 3.24  &$8.99 \times 10^{7}$ \\
         1.75 & $8.30 \times 10^{6}$ & 0.679 & 0.605 & 57.6 &    $6.25  \times 10^{8}$   & 3.69&$2.87 \times 10^{8}$ \\
         0.392 & $7.86 \times 10^{6}$ & 0.258 & 0.647 & 21.5 &   $6.80  \times 10^{7}$   & 1.43& $1.06 \times 10^{9}$ \\
         \hline
    \end{tabular}
\end{table*}

\begin{table*}
	\centering
	\caption{This table shows 
	the timescales (in years) of each portion of the orbital decay for all five resolvable mergers 
	at their respective redshifts. 
	For the two main portions of the coalescence (orbital decay and hardening), we take the minimum timescale calculated due to stars or gas, and use that time to calculate the overall delay timescale, which is the sum of dynamical friction and hardening timescales.    The coalescence time is the age of the universe when black holes merge plus the delay timescale.}
	\label{tab:all}
	\begin{tabular}{|c|p{15mm}|c|p{15mm}|p{15mm}|p{15mm}|p{15mm}}
		\hline
		Redshift  & $T_{\rm OD,star}$ x 10& $T_{\rm OD,gas}$ & $T_{\rm h,GW}$ & $T_{\rm h,gas}$ & Delay Timescale &  Coalescence Time \\
	
		\hline 
		9.42    & $6.02 \times 10^{8}$  &    $1.40 \times 10^{6}$     &  $3.16 \times 10^{10}$ & $3.10 \times 10^{9}$  & $3.10 \times 10^{9}$ & $3.62 \times 10^{9}$    \\
		4.60        & $9.69 \times 10^{7}$  &   $3.47 \times 10^{8}$      &$3.73 \times 10^{9}$     &  $2.35 \times 10^{10}$&  $3.83 \times 10^{9}$&  $5.16 \times 10^{9}$ \\
		3.79        & $6.11 \times 10^{6}$  &   $1.69 \times 10^{11}$     & $7.84 \times 10^{8}$   & $9.60 \times 10^{8}$& $7.90 \times 10^{8}$ &  $2.47 \times 10^{9}$   \\
		1.75     & $7.18 \times 10^{7}$  &   $1.32\times 10^{8}$       & $8.01 \times 10^{9}$   &  $5.47\times 10^{10}$   &  $8.08 \times 10^{9}$ &  $1.19 \times 10^{10}$   \\
		0.392    & $1.28 \times 10^{8}$   &   $3.40 \times 10^{7}$      &$2.16 \times 10^{9}$    &  $7.28 \times 10^{10}$& $2.19 \times 10^{9}$& $1.17 \times 10^{10}$   \\
		\hline
	\end{tabular}
\end{table*}

\subsection{Gravitational Wave Signals}
\label{sec:Gravitational Wave Signals}

\begin{figure}

	\includegraphics[width=\columnwidth]{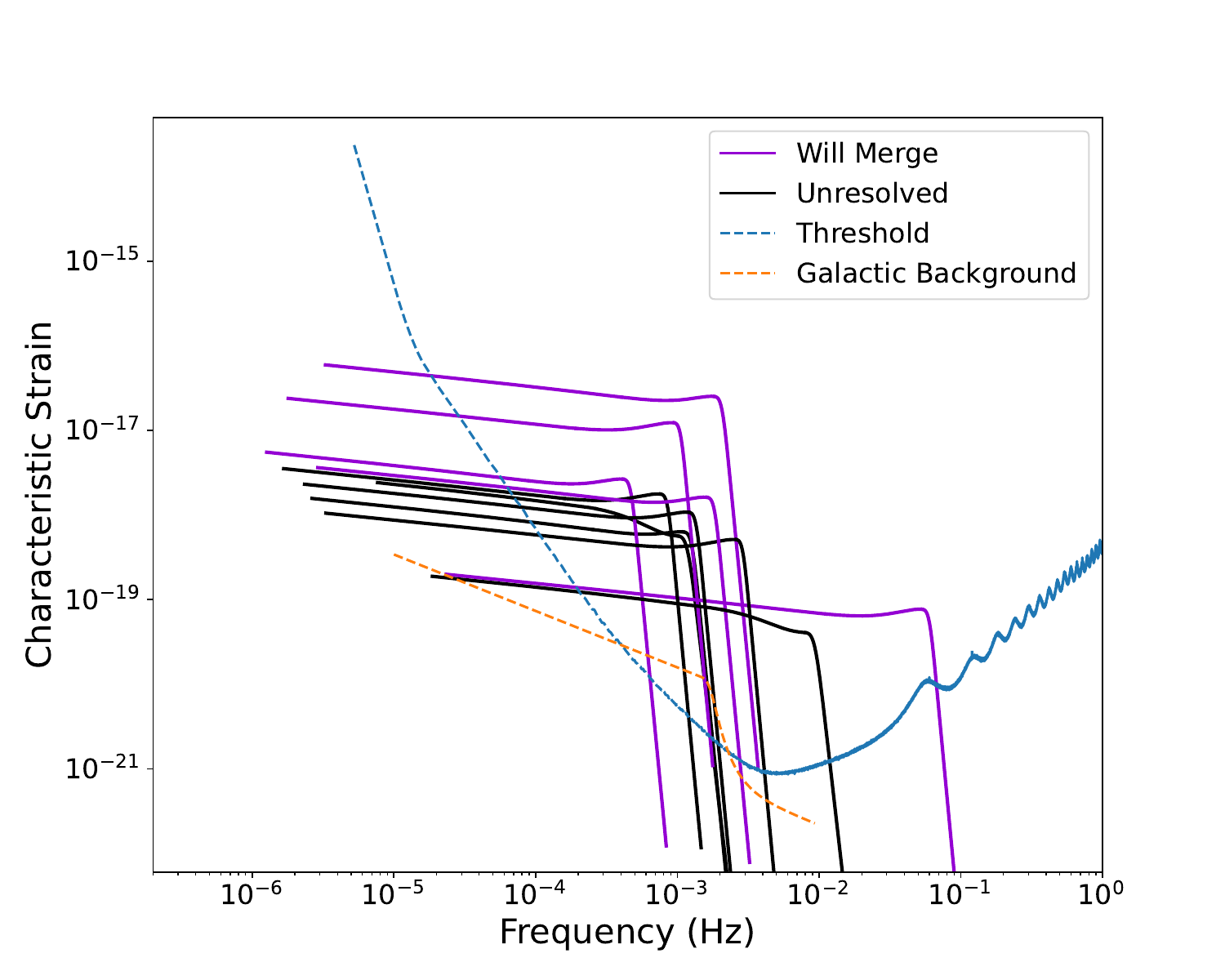}
    \caption{In this gravitational wave strain plot, the y-axis, Characteristic Strain, is used to model the binary signal which accounts for the time the binary spends in each frequency bin, and the x-axis is the frequency of gravitational waves.  We do not explicitly calculate signal-to-noise ratios here, but rather just estimate the detectability using the characteristic strain of the signal. Lines showcase the inspiral, where black holes are getting closer together until the merger, and finally the ringdown, where spacetime adjusts and removes distortions away from an axi-symmetric Kerr black hole. The lines in black are unresolved mergers, which we show for informational purposes; lines in purple show which will merge within a Hubble time. The Sensitivity Curve in dashed blue tells us that the gravitational waves above this line will allow LISA to detect these gravitational waves, and the dashed orange is the galactic background. The galactic background noise is originally suggested in \citet{BenderHils1997}, and we use the analytical approximation of the galactic background noise.} 
    \label{fig:GWplot}
\end{figure}

  We use \citet{Katz2019}'s method in order to produce the frequency-domain gravitational wave strain plot displayed in Figure \ref{fig:GWplot}, where characteristic strain is used to model the binary signal which accounts for the time the binary spends in each frequency bin (see \citet{Katz2019} for a deeper explanation). We do not explicitly calculate signal-to-noise ratios here, but rather just estimate the detectability using the characteristic strain of the signal. We produced Figure \ref{fig:GWplot} after determining the total mass, mass ratio, and redshift of binary black holes (see Table \ref{tab:galaxy properties}). It compares the 11 different binary black holes that merge inside dwarf galaxies from our simulations. This plot showcases the inspiral, where black holes are getting closer together until the merger, and finally, the ringdown where spacetime adjusts and removes distortions away from an axi-symmetric Kerr black hole. Black lines are the unresolved mergers, which we show for informational purposes. The lines in purple show which will merge within a Hubble time, while the pink line will not merge. The dashed blue line, labeled "Sensitivity Curve" tells us where LISA will be able to detect the gravitational waves, and the orange dashed line shows us the galactic background due to white dwarf binaries. The galactic background noise is originally suggested in \citet{BenderHils1997}, and we use the analytical approximation from \citet{Hiscock2000} to include the effect of the galactic background noise. All eleven close pairs in our simulation would be detectable by LISA if they indeed are able to merge efficiently.

\begin{figure}
	
	\includegraphics[width=\columnwidth]{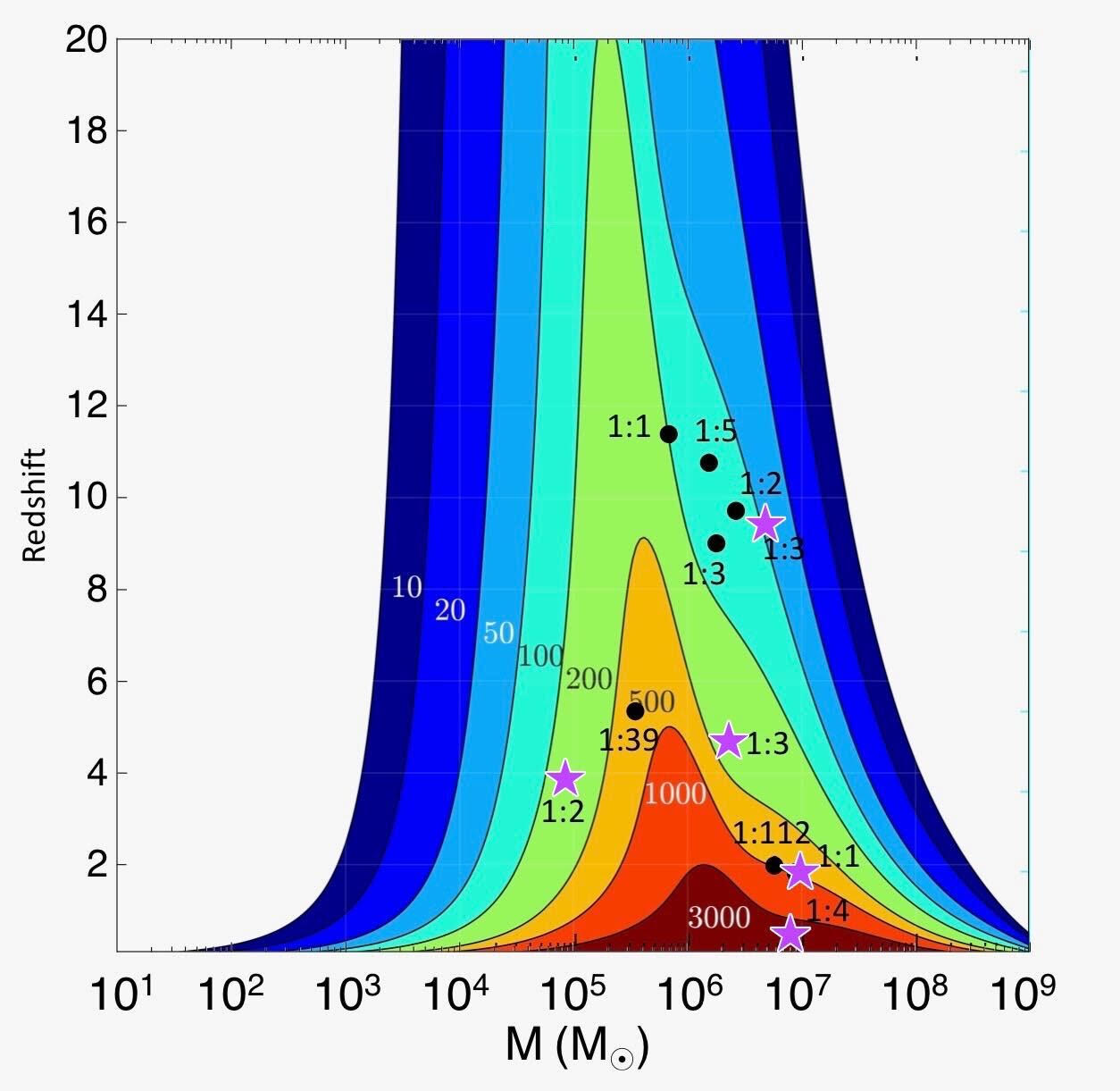}
    \caption{Originally from \citet{Bellovary2019}, this plot shows the 11 different binary black hole merger events in dwarf galaxies from the cosmological simulations. On the y-axis is redshift, and x-axis is total mass in solar masses. Points show these merger events as a function of redshift and the combined black hole mass. Near each point, is the mass ratio of each merger. Rainbow contours and smaller numbers represent the Signal-to-Noise ratio with which LISA will detect such mergers if they have a characteristic mass ratio of 1:4, with red being the most detectable, and blue being the least. This plot has been modified to show black points and purple stars, where black points are unresolved, and purple stars will merge in a Hubble time.  }
    \label{fig:Rainbow Plot}
\end{figure}

Figure \ref{fig:Rainbow Plot} (adapted from \cite{Bellovary2019}) shows the redshift vs total black hole mass for 11 binary black hole merger events in dwarf galaxies from the cosmological simulations mentioned earlier.  The points show each merger event, and near each point is the mass ratio of each merger. Rainbow contours and smaller numbers represent the Signal-to-Noise ratio with which LISA will detect such mergers, if they have a characteristic mass ratio of 1:4. This plot has been modified to show black and purple points, where black points are unresolved mergers, and purple points will merge within a Hubble time and can be detected by LISA. Each of these mergers would have a signal-to-noise of at least 100 if observed by LISA.

\subsection{Comparisons to Prior Work}
\label{sec:Comparisons to Prior Work}

Our results are in slight tension with \citet{Tamfal18}, who examine black hole pairing using $N$-body simulations of dwarf galaxy mergers.  They show that pairing and merger will be relatively swift in dwarfs with cuspy density profiles, but in dwarfs with cored profiles the formation of a hard binary takes longer than a Hubble time.  All of our dwarfs have cored density profiles (also see Figure 6 in \citet{Bellovary2021}), and in Section \ref{sec:DA17} we discussed the issues regarding sinking bodies in cored potentials.  Further work is needed  in simulating sinking black holes in cored potentials that include gas physics and realistic dynamics, which cause asymmetries and other perturbations which may accelerate dynamical friction (or do the opposite).

Recent work by \citet{Khan21} examines intermediate mass black hole mergers in dwarf galaxies using ultra high-resolution $N$-body simulations.  Using realistic models based on local dwarf galaxies, they evolve binaries through the dynamical friction regime and into the weak hardening regime.  They find that binaries shrink from 50 pc to $\sim$1 pc in 5-20 Myr, depending on the structural properties of the host galaxy;   our dynamical friction timescale estimates are consistent with these values.  They predict final coalescence times are on the order of a few 100 Myr, while our estimates are somewhat longer.  In the event that we have overestimated hardening and coalescence times, our final measurements are conservative, and mergers in dwarf galaxies could happen  more quickly than we predict. 

This topic has also been explored analytically in the context of stellar black holes merging in ultrafaint dwarf galaxies, in order to explain recent results from LIGO/Virgo.  \citet{Conselice20} argue that black holes with masses of $10-80 M_\odot$ could merge within the remnants of merged dwarf galaxies within a Hubble time, and that this rate may be sufficient to explain many of the existing gravitational wave detections, including the extremely massive GW190521 \citep{Palmese21}.  These objects are much less massive than those we discuss here, and it is unclear if their sinking timescales would truly be as rapid as suggested.  Overall, further study  on this topic is needed to solve the problem of black holes sinking to the centers of cored potentials.

\section{Conclusions}

Previous cosmological hydrodynamic simulations have shown the prediction of massive black holes merging in dwarf galaxies, but these simulations are limited by their resolution and cannot follow black hole pairs all the way to coalescence. They do not include important physics such as dynamical friction due to gas or stars, loss-cone scattering, or hardening due to gravitational radiation, which are all necessary in order to calculate realistic delay times for black hole pairing and mergers. This research provides the calculation of delay timescales based on the properties of the black holes and the dwarf galaxies, which does account for dynamical friction due to gas and stars, loss-cone scattering, and hardening of the binary due to gravitational radiation and gas dynamics.

We have calculated delay timescales for black hole mergers which take place throughout cosmic time, and the results are summarized below: 
\begin{itemize}
    \item Out of the 11 mergers, five have coalescence times shorter than a Hubble time (between 0.8 - 8 Gyr), and six do not have sufficient resolution to determine merger characteristics.
    \item  All 11 mergers have  characteristic strains that are detectable by LISA. 
    \item As all five of the resolved close pairs merge within a Hubble time, we make the broad generalization that in the event that two massive black holes exist within a low-mass galaxy, they are likely to find each other and merge within $\sim 3$ Gyr.    
\end{itemize}

Black holes merging in dwarf galaxies may be a common phenomena that LISA will be able to detect, thus it is important to continue this research in order to prove this and provide information regarding these black hole merger events in dwarf galaxies. The next logical step is measuring a black hole merger rate in dwarf galaxies over cosmic time, which requires knowledge of the occupation fraction of MBHs in dwarfs as a function of both mass and redshift, as well as a much larger sample of simulated galaxies than we present here.  In the future we plan to repeat these calculations for the ROMULUS simulation \citep{Tremmel17}, which will allow us to more robustly determine the fraction of close black hole pairs that merge, as well as calculate global merger rates. Our simulation has 165 galaxies, but the ROMULUS simulation has $\sim 1000$ galaxies ranging from dwarf galaxies to massive galaxies. Having a larger sample size from ROMULUS will contribute to making a stronger conclusion regarding the coalescence and rates of merging MBHs. Our current analysis gives us an initial idea that black hole mergers in dwarf galaxies could be a substantial contribution to LISA's detected signal, indicating future work on this topic will be of great interest to the gravitational wave community. 


\section*{Acknowledgements}

We are grateful to the anonymous referee for constructive feedback which greatly improved the depth of the paper.  VIDC is grateful to the National Science Foundation's Research Experiences for Undergraduates program hosted by the American Museum of Natural History (AST-1852355), and to Ray Sharma for providing helpful comments.  JMB acknowledges support from NSF AST-1812642 and the CUNY JFRASE award.   

\section*{Data Availability}

The simulations analyzed in this work (DC Justice League and MARVEL-ous Dwarfs) are proprietary and are not available to the public.  The authors are happy to share quantitative data related to our results  for collaborative purposes upon request.



\bibliographystyle{mnras}





\bsp	
\label{lastpage}
\end{document}